\documentclass[twocolumn,twoside,slac_two]{revtex4}
\usepackage{graphicx}
\usepackage{fancyhdr}
\usepackage{amssymb,amsmath}
\newcommand{\LNP}{\Lambda_{\text{NP}}}
\begin{document}

\title{Heavy Flavor Theory}

%

\author{B. Grinstein}
\affiliation{Physics Department, University of California, San Diego,
  La Jolla CA 92093-0319, USA}

\begin{abstract}
  This is a limited review and update of the status of Heavy Flavor
  Physics. After we review the flavor problem we discuss a number of
  topics: recent puzzles in purely leptonic $D$ and $B$ decays and
  their possible resolutions, mixing in neutral  $B$ and $D$ mesons,
  the determination of $|V_{cb}|$ and $|V_{ub}|$ from semileptonic
  decays, and we conclude with radiative $B$ decays.  
  
\end{abstract}

\maketitle

\thispagestyle{fancy}


\section{Introduction: The Flavor Problem}
The Standard Model (SM) of electroweak
interactions\footnote{Supplemented with right handed neutrinos and masses.} correctly
accounts for all known particle physics data. Hints of small anomalies
exits but none is firmly established. If there is new physics (NP) it must
be hiding, and one good way to hide it is by making it active only at
shorter distances than we have yet probed. Future high energy particle
collision experiments may directly probe such new short distance
physics. We can ask, in the mean time, what are the indirect effects
of such new physics at the longer distances that are probed in current
experiments? A model independent way to address this question is by
supplementing the Lagrangian of the SM with local terms, or ``operators,''
of dimension greater than four. Such terms render the theory
non-renormalizable. Hence a momentum cut-off $\LNP$ must be introduced
and the theory is considered only as an Effective Field Theory (EFT) valid only at
energies below $\LNP$. Colliding particles with center of mass energy
in excess of $\LNP$ surely produces new states that require further
specific modification of the theory.  

The scale $\LNP$ also serves to make the terms in the Lagrangian
dimensionally correct. An operator of dimension $n>4$ appears with
coefficient $c/\LNP^{n-4}$, with $c$ a dimensionless constant. The
natural expectation is that $c$ is of order unity. A very large number
renders the theory in-effective, breaking down at energies below
$\LNP$. This does not happen: one simply chooses a smaller number for
$\LNP$. On the other hand, the coefficients of some  operators could be
very small. But short of explaining why some coefficients unexpectedly
small, we must assume that in fact we underestimated $\LNP$. Hence, we
may proceed by assuming $c\sim 1$ and see what current data implies
for $\LNP$. 

The EFT generically contains $\Delta F=2$ FCNCs, that is, terms that
induce neutral interactions that change flavor by two units. For
example one may include
\begin{equation}
\label{eq:bounds}
\frac1{\LNP^2}\left[c_1(\bar d_L\gamma^\mu s_L)(\bar d_L\gamma_\mu s_L)
+c_2(\bar u_L\gamma^\mu c_L)(\bar u_L\gamma_\mu c_L)\right].
\end{equation}
Then, if one ignores the SM contribution, neutral meson mixing data gives\cite{Blum:2009sk}
\begin{align}
c^{(\text{data})}_1&=(8.8+0.033i)\times10^{-7}\left(\frac{\LNP}{\text{1
      TeV}}\right)^2,\\
\label{eq:DDEFTbound}
c^{(\text{data})}_2&=(5.9+1.0i)\times10^{-7}\left(\frac{\LNP}{\text{1
      TeV}}\right)^2.
\end{align}
There being no reason to expect a cancellation between the SM
and NP contributions, the NP contributions should  be smaller than
$c_i^{(\text{data})}$. Therefore $\LNP$ should be larger than the
electroweak scale by some four orders of magnitude!

This in itself is not a problem. But there is one good reason to
expect NP at the electroweak scale. In the SM there are quadratically
divergent radiative corrections to the higgs mass. In terms of our
cut-off EFT, the  shift in the higgs mass from  $L$-loops is of order
$\LNP/(4\pi)^L$ so a counterterm must be fine tuned to one part per
mil to cancel this at one loop, and further fine tuned to one per cent
at two loops, etc. If instead $\LNP\sim1$~TeV there is no need for any
fine tuning. This is the Flavor Problem, that NP at the EW scale
requires extraordinarily small dimensionless couplings $c_i$. 

Much of the work on Heavy flavor physics aims at testing the
SM in the flavor sector with high precision. It gives additional
restrictions on NP, that can be described as bounds on additional
coefficients of higher dimension operators. The purpose of this talk
is to present some of the main results in heavy flavor theory. While
it is interesting to investigate models of NP that address the Flavor
Problem, a prerequisite is to understand the restrictions that heavy
flavors place on the models. In the absence of glaring anomalies in
the data, this is best done by verifying consistency of the SM to as
high precision as possible. I will focus on precision SM
determinations, but will here and there indicate implications on  models of NP.

\section{Purely Leptonic Decays}
\subsection{The evanescent $\mathbf{f_{D_s}}$ puzzle}
The theory of purely leptonic decays is simple,
\begin{equation}
\Gamma(D_s\to \ell\nu_\ell)
=\frac{m_{D_s}}{8\pi}f_{D_s}^2G_F^2m_\ell^2|V_{cs}|^2(1-m_\ell^2/m_{D_s}^2)^2
\end{equation}
with $f_{D_s}$ the $D_s$ decay constant and $V$ the Kobayashi-Maskawa
matrix. There are analogous formulas with obvious modifications when
replacing $B$, $B_s$ or $D$ for $D_s$. Last year a discrepancy became
apparent in the value of $f_{D_s}$ obtained from Monet-Carlo
simulations of QCD on the lattice and the one from the experimental
measurement of the purely leptonic branching fraction. From a recent
compilation of lattice results\cite{Onogi:2009vc}
\begin{equation}
f_D=206(4)~\text{MeV},\qquad f_{D_s}=243(3)~\text{MeV}
\end{equation}
while experimentally\cite{:2008sq,:2007ws}
\begin{equation}
f_D=205.8(8.5)(2.5)~\text{MeV},\quad f_{D_s}=275(16)(12)~\text{MeV}
\end{equation}
While this anomaly was not firmly established, the agreement between
lattice and experiment in the value of $f_D$ suggests the discrepancy
in $f_{D_s}$ may well remain once the errors are reduced. 

Perhaps for this reason several groups have looked for a viable
interpretation of this result in terms of
NP\cite{Dobrescu:2008er,Benbrik:2008ik,Dorsner:2009cu}. In the SM this
is a Cabibbo allowed, tree level decay. Hence for the NP to have a
significant effect it must be neither loop nor Cabibbo
suppressed. Moreover the mass $M$ of the new particle mediating this
interaction cannot be too large: for constructive interference the
amplitude should be about 6\% of the SM's, so roughly $M\approx
M/\sqrt{0.06}=320$~GeV.  Dobrescu and Kleinfeld argued that (i)
$s$-channel charged higgs exchange could explain the effect, with
$y_s\ll y_c$ and both $y_c$ and $y_\tau$ of order unity, but then
found this explanation disfavoured by $D$ decay data (ii) $t$-channel
charge +2/3 leptoquark exchange could also account for the data but is
disfavored by the bound on $\tau\to\bar s s\mu$ (iii) $u$-channel charge
$-1/3$ leptoquark exchange ($d$-squark like object, $\tilde d$) is a viable
explanation. They introduce the interaction
\begin{equation}
{\cal L}_{\text{LQ}}= \kappa_\ell(\bar c_L\ell^c_L-\bar s_L\nu^c_{\ell
  L})\tilde d+\kappa'_\ell\bar c_R\ell^c_R\tilde d +{\text{h.c.}},
\end{equation}
which is already present in supersymmetric extensions of the SM
without R parity, and show that for $|\kappa'_\ell/\kappa_\ell|\ll
m_\ell m_c/m^2_{D_s}$ the resulting interference is automatically
constructive. In addition, if $|\kappa_\ell|\approx|\kappa_\tau|$  the
deviations in $\mu\nu$ and $\tau\nu$ are approximately equal. 

Earlier this year the CLEO collaboration published new results on
$D_s$ purely leptonic decays to both  $\mu\nu$ and $\tau\nu$ final
states\cite{Alexander:2009ux}. Their value for the decays constant,
$f_{D_s}=259.5(6.6)(3.1)$~MeV has significantly reduced errors but
also has moved significantly in the direction of eliminating the
anomaly.

\subsection{$\mathbf{B\to\tau\nu}$}
\label{btotaunu}
\begin{figure}
\includegraphics[width=80mm]{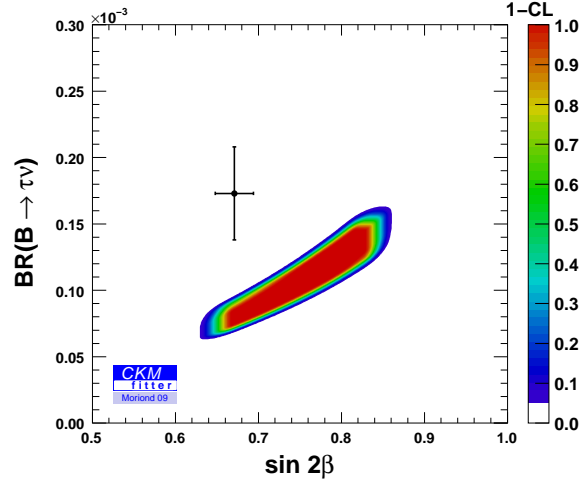}%
\caption{\label{fig:tension_Btau} Tension in the CKM global fit from
  $B\to\tau\nu$\cite{CKMfitter}. The cross corresponds to the
  experimental values with 1 sigma uncertainties. The shaded region is
  the result of the global fit performed without using these
  measurements.  }
\end{figure}

The direct determination of the branching fraction for $B\to\tau\nu$ and
$\sin2\beta$, which measures the CP asymmetry in interference
between mixing and decay in $B\to J/\psi K_s$, is in slight
disagreement with  a global fit to these quantities from other
measurements. This is presented in Fig.~\ref{fig:tension_Btau} which
shows both the result of the direct measurements (as a cross) and that
of the global fit (as a shaded area). 

That there is a correlation between these quantities can be seen from
the following formula, obtained from the expression for the $B^0-\bar
B^0$ mass difference\footnote{See next section.} $\Delta m_d$, by
eliminating the decay constant $f_B$ in favor of the purely leptonic
branching fraction:
\begin{equation}
\label{BrtoDeltaMratio}
\frac{\text{Br}(B\to\tau\nu)}{\Delta m_d}
= \frac{3\pi m_\tau^2(1-\frac{m_\tau^2}{m_B^2})^2\tau_{B^+}}{4m_W^2S(x_t)|V_{ud}|^2}
\frac{1}{B_{B_d}}\left(\frac{\sin\beta}{\sin\gamma}\right)^2.
\end{equation}
Here $S(x_t)$ is an Inami-Lim function with the top quark mass as
argument, and $B_{B_d}$ is the ``bag'' constant that parametrizes the
matrix element of the short distance four-quark operator that gives
rise to mixing. The main uncertainties come from $B_{B_d}$ and the angles
$\alpha$ and $\gamma$. 

The deviation is 2.4 sigmas if one compares the indirect fit prediction
for the $B\to\tau\nu$ branching fraction with the direct measurement.
Alternatively one can use the fit to determine $B_{B_d}$ and compare
with the value determined form Monte-Carlo simulations of QCD on the
lattice. Then the deviation is 2.7 sigmas.

\begin{figure}
\includegraphics[width=80mm]{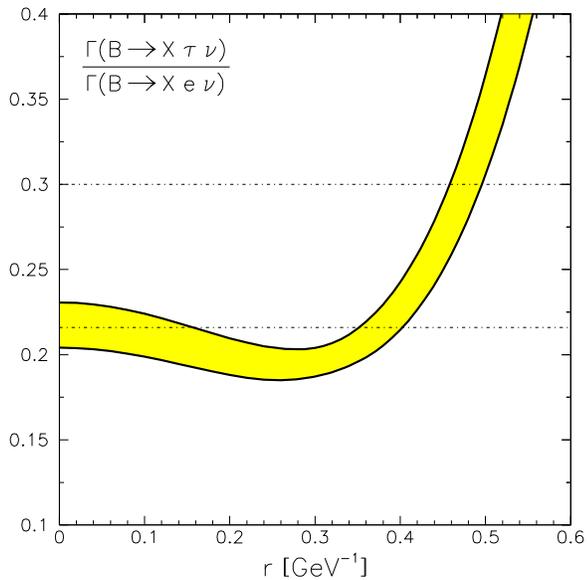}%
\caption{\label{fig:BtauvsE} The shaded area gives the predicted ratio
  of $B$ semileptonic branching fractions into $\tau$ vs. electron, as
  a function of $r=\tan\beta/M_{H^+}$, in a two higgs doublet
  model\cite{Grossman:1995yp}.  The dashed-dotted lines give
  experimental 1$\sigma$ bounds.  }
\end{figure}

As in the $D_s$ decay case this is a tree level process, but in
contrast, this is Cabibbo suppressed, which gives a bit more leeway in
giving a NP interpretation. In a two higgs doublet model tree level
charged higgs exchange can easily modify the leptonic branching
fraction\cite{Hou:1992sy,Isidori:2001fv}. In this model the Cabibbo
suppression is still present but there is an amplification factor
proportional to $\tan^2\beta=(v_2/v_1)^2$, where $v_1$ and $v_2$ are
the vacuum expectation values of the higgs doublets responsible for up
and down masses, respectively. The parameter that controls the NP
correction is $r=\tan\beta/M_{H^+}$. The semileptonic branching
fraction will also be affected, and a sensitive test of this
interpretation is the ratio of semileptonic decay branching fractions
into $\tau$ versus electrons\cite{Grossman:1995yp}; see Fig.~\ref{fig:BtauvsE}.

\section{Neutral Meson Mixing}
\subsection{Generalities}
The recent measurements of mixing of neutral $D$ mesons and their
unfamiliar properties suggest we begin our discussion by a taking a
general look at neutral meson mixing. Let us briefly review the
quantities that enter the description of neutral meson
mixing. Parameters $p$ and $q$ are introduced to express the physical
states in terms of flavor eigenstates: $|P_{L,H}\rangle = p
|P^0\rangle\pm q|\bar P^0\rangle$. These, together with the mass and
width differences, $\Delta m= m_H - m_L$ and $\Delta \Gamma =
\Gamma_H-\Gamma_L$, are given in terms of the off-diagonal elements of
the $2\times2$ non-hermitian hamiltonian $M-\frac{i}2\Gamma$ by
\begin{equation}
\label{deltadelta}
\begin{aligned}
(\Delta m)^2-{\textstyle\frac14}(\Delta\Gamma)^2 &=
4|M_{12}|^2-|\Gamma_{12}|^2,\\
\Delta m\Delta\Gamma &= 4\text{Re}\,(M_{12}\Gamma^*_{12}),\\
\frac{q^2}{p^2} &=
\frac{2M^*_{12}-i\Gamma^*_{12}}{2M_{12}-i\Gamma_{12}}.
\end{aligned}
\end{equation}
We expect $M_{12}$ rather than $\Gamma_{12}$ to be more prone to
modifications from new physics because $\Gamma$ is largely given by
long distance physics. Finally, the decay amplitudes are 
\begin{equation}
A_f=\langle f|{\cal H}|P^0\rangle\quad\text{and}\quad \bar A_f=\langle f|{\cal H}|\bar P^0\rangle.
\end{equation}

CP violation (CPV) can be searched through processes that probe different
quantities. If $|\bar A_f/A_f|\ne1$ there is CPV in decays, while
$|q/p|\ne1$ gives CPV in mixing. A non-vanishing imaginary part of
$\lambda_f=(q/p)(\bar A_f/A_f)$ gives CPV in the interference between
mixing and decay. The phase
$\phi_{12}=\text{arg}(-M_{12}/\Gamma_{12})$ is sensitive to NP that
may show up in $M_{12}$. The parameter that controls the di-lepton
asymmetry is $\text{Im}\Gamma_{12}/M_{12}=(1-|q/p|^4)/(1+|q/p|^4)$; it
is non-perturbative and hence difficult to compute (the OPE is no
better that for lifetimes, perhaps worse). 

The behavior of the mixing system depends rather sensitively on which
of $\Delta m$ and $\Delta \Gamma$ is largest. Consider first the case
$\Delta m\gg\Delta \Gamma$ which is the situation for $B$ and $B_s$
mesons. This condition corresponds to small $\Gamma_{12}/M_{12}$ and
one can find approximate solutions of \eqref{deltadelta}: $\Delta
m=2|M_{12}|(1+\cdots)$ and
$\Delta\Gamma=-2\Gamma_{12}\cos\phi_{12}(1+\cdots)$, where the
ellipsis indicate corrections of order $\Gamma_{12}/M_{12}$. Keeping
in mind that $\phi_{12}$ is suppressed in the SM, we see that the
effects of NP can only reduce the magnitude of $\Delta\Gamma$. On the
other hand since  $q/p=-\text{arg}(M_{12})(1+\cdots)$  time dependent CP
asymmetries are sensitive to NP that may show up in $M_{12}$. 

The other extreme case has $\Delta \Gamma\gg \Delta m$ and this
condition corresponds to small $M_{12}/\Gamma_{12}$. As before, one
can find approximate solutions of \eqref{deltadelta}: $\Delta m =
2|M_{12}\cos\phi_{12}|(1+\cdots)$,
$\Delta\Gamma=\mp2\Gamma_{12}(1+\cdots)$ and
$q/p=-\text{arg}(\Gamma_{12})(1+\cdots)$ depends weakly on $M_{12}$
(the ellipsis now indicate corrections of order
$M_{12}/\Gamma_{12}$). For example, if $D$-mesons satisfy $\Delta
\Gamma\gg \Delta m$ and there in negligible CPV in the decay, then
$\text{arg}\lambda_{K^+K^-}\propto|M_{12}/\Gamma_{12}|^2\sin(2\phi_{12})$.
Hence there is reduced sensitivity to NP in $M_{12}$, even for
dominant NP.

\subsection{$\mathbf{B^0\bar B^0}$ and $\mathbf{B_s\bar B_s}$}
CPV in $B$ decays and mixing is discussed at length in other talks at
this conference. In order to avoid unnecessary duplication we limit
ourselves to discussing the mass difference measurements and theory. 

In the SM $B^0$ and $B_s$ mixing have the same underlying dynamics
(double $W$ exchange, with virtual top-antitop quark intermediate
state). Hence the expressions\footnote{Which are rather involved,
  witness Eq.~\eqref{BrtoDeltaMratio}.} for $\Delta m$ are virtually
identical, except for obvious change of parameters. Taking the ratio
not just simplifies the expressions but also cancels some
uncertainties. Solving for the ratio of KM elements
\begin{equation}
\frac{|V_{ts}|}{|V_{td}|}=\xi\sqrt{\frac{\Delta m_s \; m_{B_s}}{\Delta
    m_d \; m_{B_d}}}, \quad\text{where}\quad
\xi^2=\frac{B_{B_s}f^2_{B_s}}{B_{B_d}f^2_{B_d}}.
\end{equation}
Here $f_B$ and $B_B$ are the decay and bag constants (see
\eqref{BrtoDeltaMratio}) introduced in previous sections. The quantity
$\xi$, required to extract a value for the ratio of KM elements,
contains all of the hard to estimate hadronic physics and has the
property that it is unity in the $SU(2)_v$ symmetry limit
($m_d=m_s$). Monte-Carlo simulations of lattice QCD give
\begin{equation}
\begin{aligned}
\xi&=1.205(52)\qquad\text{FNAL/MILC\cite{ToddEvans:2008}},\\
\xi&=1.258(25)(21)\qquad\text{HPQCD\cite{Gamiz:2009ku}}.
\end{aligned}
\end{equation}

Using Belle and BaBar accurate measurements of $\Delta m_d$ and CDF
and D0 measurements of $\Delta m_s$, Evans reports\cite{Evans:2009xf}
\begin{equation}
\frac{|V_{td}|}{|V_{ts}|}=0.2060\pm0.0012(\text{exp})^{+0.0081}_{-0.0060}(\text{theor}).
\end{equation}
I hasten to point out that Evans uses a value for $\xi$ reported in
2003\cite{Aoki:2003xb}, one of the early unquenched
calculations. Luckily, this value interpolates between the
two more recent results, $\xi=1.210(^{+47}_{-35})$, and the errors are
also similar. Rather than fussing over the best central
value and theoretical error let us stop to think how well we can trust
lattice calculations with this extraordinary, $\sim3\%$ precision. In
fact, the precision of the calculation is only about  16\% since only
the deviation of $\xi^2$ from unity needs be computed. Some other
rather crude methods should therefore work rather well too. For
example, a very early computation of the chiral logs gives
$\xi=1.14$\cite{Grinstein:1992qt}. If we estimate the errors by
$\Delta\xi^2\sim(m_K/\Lambda_\chi)^2\approx20\%$, we would write
$\xi=1.14\pm0.08$, or a 7\% error. Remarkably, this crude
determination agrees with the lattice result, within expected errors!

\begin{figure}
\includegraphics[width=80mm]{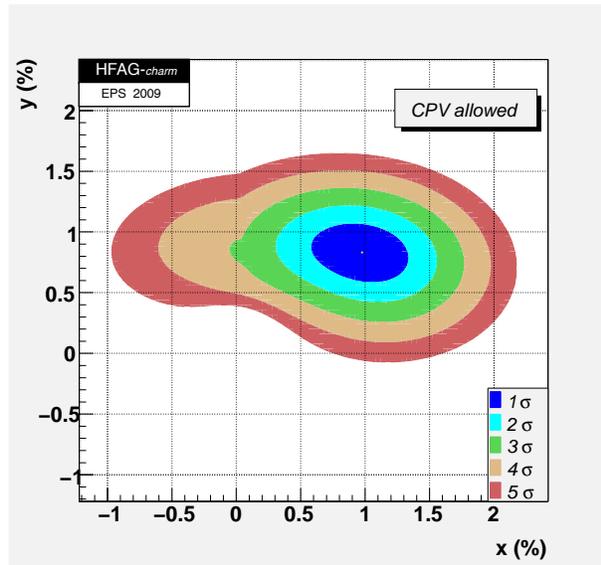}%
\caption{\label{fig:DDosc} Contours of allowed values of the
  parameters $x=\Delta m/\Gamma$ and $y=\Delta
\Gamma/\Gamma$ from an HFAG analysis of the Belle and BaBar data on
$D^0$ mixing and decays\cite{HFAG-DDbar-CPV}.   }
\end{figure}

\subsection{$\mathbf{D^0\bar D^0}$}
In the two years that followed the first evidence for $D^0\bar D^0$
oscillation by BaBar\cite{Aubert:2007wf} and
Belle\cite{Staric:2007dt}, these two collaborations have reported a
number of related and improved
measurements\cite{Aubert:2007aa,Aubert:2007en,Aubert:2008zh,Aubert:2009ck,Abe:2007rd,Bitenc:2008bk,Zupanc:2009sy}. It
is customary to introduce $x=\Delta m/\Gamma$ and $y=\Delta
\Gamma/\Gamma$.  The experimental situation is summarized in
Fig.~\ref{fig:DDosc}, which shows the allowed region in the $x,y$
plane.

The two-$W$ exchange graph that gives $\Delta F=2$ processes in the SM
has, in the $D\bar D$ case, an intermediate $q\bar q'$ state with
$q,q'=d,s,b$. Since these are light compared to the $W$, GIM
suppression is very effective in $x$ and~$y$. Therefore $x,y$ are
small. But they are not perturbatively calculable.  They are probably
dominated by the same long distance physics so $x$ may well be
comparable in size to $y$. Moreover, $y$ has a Cabibbo suppression
factor $\sin^2\theta_C$ and vanishes in the $SU(2)_v$ limit
($m_s=m_d$).  

The experimental measurement $x\approx y\approx1\%$ is very compatible
with SM expectations. However, since we can't compute accurately,
precision tests of the SM are not possible. Still one can use the
measured values of $x$ and $y$ to constrain models of NP. $M_{12}$ is
more sensitive to NP because, in the SM, it starts at 1-loop while
$\Gamma_{12}$ starts at tree level. Moreover, NP is short distance
dominated. Excluding a possible perverse cancellation between the
short and long distance contributions to $M_{12}$ we can then restrict
the NP by demanding that its contribution be no larger than what is
measured. We have already alluded to this in a model independent way
in \eqref{eq:DDEFTbound}. The implications for many specific models
have been studied, including the MSSM and a sequential fourth
generation of quarks \cite{Ciuchini:2007cw,Golowich:2007ka}. The MSSM
can evade the bound by the same mechanisms that are available for the
other flavor problems of that model: either make the SUSY breaking
scale uncomfortably high or find a mechanism, like gauge mediation,
that implements Minimal Flavor Violation in the MSSM. Because it  must
 be very heavy, a fourth
generation quark in the two-$W$ exchange graph does not have an
effective GIM mechanism. But its contribution can be suppressed by
small  KM ($4\times4$) matrix elements, $|V_{ub'}V_{cb'}|\sim0.001$
for $m_{b'}\sim$ few 100's GeV.

\section{Determination of $\mathbf{|V_{cb}|}$}
\label{sec:vcb}
This section and the next are concerned with the precision determination
of the magnitudes of the two KM elements $V_{cb}$ and $V_{ub}$. It
goes without saying that a precise determination is fundamental in
testing the KM theory of  CP violation in the SM. Traditionally these
magnitudes are determined from the semileptonic decay rate of $B$
mesons. Hence these sections are largely about the theory of
semileptonic $B$ decay, and we separate each one into inclusive and
exclusive sub-sections since the theory is different.

\subsection{Inclusive}
The theory of inclusive semileptonic $B$ decay is based on a double
expansion, combining the heavy mass expansion of Heavy Quark Effective
Theory (HQET) with the Operator Product Expansion
(OPE)\cite{Chay:1990da,Bigi:1993fe,Manohar:1993qn}. The decay rate can
then be expressed as an expansion, $\Gamma=\Gamma_0+\Gamma_1+\cdots$,
where $\Gamma_n\sim(\Lambda_{\text{QCD}}/m_b)^n$. The main results are
that $\Gamma_0$ is actually given by the perturbative expression for
the semileptonic decay rate for an unconfined $b$ quark and that
$\Gamma_1=0$.

\begin{figure}
\includegraphics[width=80mm]{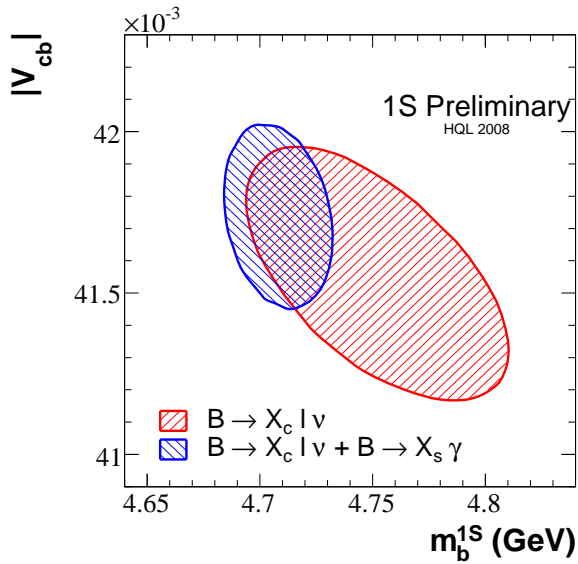}%
\caption{\label{fig:momentsResult} Results of the ﬁt combining all
  available moment data, from Ref.~\cite{Schwanda:2009bw}. The
  $\Delta\chi = 1$ contours are shown for the ﬁt with and without
  $B\to X_s\gamma$ data.  The analysis is performed assuming a 1S
  (bottom) subtraction scheme.
 }
\end{figure}

The method of moments gives a very accurate determination of
$|V_{cb}|$ from inclusive semileptonic $B$ decays. In QCD, the rate
${\rm d}\Gamma(B\to X_c\ell\nu)/{\rm d}x\,{\rm d}y=|V_{cb}|^2f(x,y)$,
where $x$ and $y$ are the invariant lepton pair mass and energy in
units of $m_B$, is given in terms of four parameters: $|V_{cb}|$,
$\alpha_s$, $m_c$ and $m_b$. $|V_{cb}|$, which is what we are after,
drops out of normalized moments, $\langle z^n\rangle=\int dx\, dy
f(x,y) z^n/\int dx\, dy f(x,y)$ where $z=x$ or $y$. Since $\alpha_s$
is well known, the idea is to fix $m_c$ and $m_b$ from normalized
moments and then use them to compute the normalization, hence
determining $|V_{cb}|$. In reality we cannot solve QCD to give the
moments in terms of $m_c$ and $m_b$, but we can use the combined
HQET/OPE to write the moments in terms of $m_c$, $m_b$ and a few
constants that parametrize our
ignorance\cite{Falk:1995kn,Kapustin:1995nr,Bauer:2002kk,Benson:2003kp,Trott:2004xc,Gambino:2004qm}. These
constants are in fact matrix elements of operators in the HQET/OPE. If
terms of order $1/m_Q^3$ are retained in the expansion one needs to
introduce five such constants; and an additional two are determined by
meson masses. All five constants and two quark masses can be
over-determined from a few normalized moments that are functions of
$E_{\rm cut}$, the lowest limit of the lepton energy integration. The
error in the determination of $|V_{cb}|$ is a remarkably small
2\%\cite{Bauer:2004ve}; see Fig.~\ref{fig:momentsResult}. But even
most remarkable is that this estimate for the error is truly
believable. It is obtained by assigning the last term {\it retained}
in the expansion to the error, as opposed to the less conservative
guessing of the next order {\it not} kept in the expansion. Since
there is also a perturbative expansion, the assigned error is the
combination of the last term kept in all expansions, of order
$\beta_0\alpha_s^2$, $\alpha_s\Lambda_{\rm QCD}/m_b$ and
$(\Lambda_{\rm QCD}/m_b)^3$.

There is only one assumption in the calculation that is not fully
justified from first principles. The moment integrals can be computed
perturbatively (in the $1/m_Q$ expansion) only because the integral
can be turned into a contour over  complex energy  $E$ away from the physical
region\cite{Chay:1990da}. However, the contour is pinned at the
minimal energy, $E_{\rm cut}$, on the real axis, right on the physical
cut. So there is a small region of integration where quark-hadron
duality cannot be justified and has to be invoked.  Parametrically
this region of integration is small, a fraction of order $\Lambda/m_Q$ of
the total. But this is a disaster because this is parametrically much
larger than the claimed error of order $(\Lambda/m_Q)^3$. However, this is
believed not to be a problem. For one thing, the fits to moments as
functions of $E_{\rm cut}$ are extremely good: the system is
over-constrained and these internal checks work. And for another, it
has been shown\cite{Boyd:1995ht} that duality works exactly in the
Shifman-Voloshin (small velocity) limit, to order $1/m_Q^2$. It seems
unlikely that the violation to local quark-hadron duality
mainly changes the normalization and has mild dependence on $E_{\rm
cut}$, and that this effect only shows up away from the SV limit.

\subsection{Exclusive}
The exclusive determination of $|V_{cb}|$ is in pretty good shape
theoretically, and only last year has become competitive with the
inclusive one. So it provides a sanity check, but not an
improvement. The semileptonic rates into either $D$ or $D^*$ are
parametrized by functions ${\cal F}$, ${\cal F}_*$, of the rapidity of
the charmed meson in the $B$ rest-frame, $w$. Luke's
theorem\cite{Luke:1990eg} states ${\cal F}={\cal F}_*=1+{\cal
  O}(\Lambda_{\rm QCD}/m_c)^2$ at $w=1$. The rate is measured at $w>1$
and extrapolated to $w=1$. The extrapolation is made with a first
principles calculation to avoid introducing extraneous
errors\cite{Boyd:1997kz}. The resulting determination of $|V_{cb}|$
has a 4\% error equally shared by theory and experiment; the theory
error is dominated by the uncertainty in the determination of ${\cal
  F}$, ${\cal F}_*$ at $w=1$. FNAL/MILC combines the 2008 PDG average
for $|V_{cb}|{\cal F}_*(1)$ with their computed value ${\cal
  F}_*(1)=0.921(13)(20)$ to obtain, after applying a small
electromagnetic correction, $|V_{cb}|=
(38.7\pm0.9_{\text{exp}}\pm1.0_{\text{theor}})\times10^{-3}$\cite{Bernard:2008dn}.

There is some tension between theory and experiment in these exclusive
decays that needs attention. The ratios of form factors $R_{1,2}$ are
at variance from theory by three and two sigma
respectively\cite{Aubert:2006cx}.  Also, in the heavy quark limit the
slopes $\rho^2$ of ${\cal F}$ and ${\cal F}_*$ should be equal. One
can estimate symmetry violations and obtains\cite{Grinstein:2001yg}
$\rho^2_{{\cal F}}-\rho^2_{{\cal F}_*}\simeq 0.19$, while
experimentally this is $-0.22\pm0.20$, a deviation in the opposite
direction. This is a good place for the lattice to make post-dictions
at the few percent error level that may lend it some credibility in
other areas where it is needed to determine a fundamental parameter.

\section{Determination of $\mathbf{|V_{ub}|}$}
The magnitude $|V_{ub}|$ determines the rate for $B\to X_u\ell\nu$. The well
known experimental difficulty is that since $|V_{ub}|\ll|V_{cb}|$ the
semileptonic decay rate is dominated by charmed final states. To
measure a signal it is necessary to either look at exclusive final
states or  suppress charm kinematically.  The interpretation of the
measurement requires, in the exclusive case, knowledge of hadronic
matrix elements parametrized in terms of form-factors, and for
inclusive decays, understanding of the effect of the kinematic cuts on
the the perturbative expansion and quark-hadron duality.

\begin{figure*}
\centering
\hspace{-1cm}\includegraphics[width=86mm]{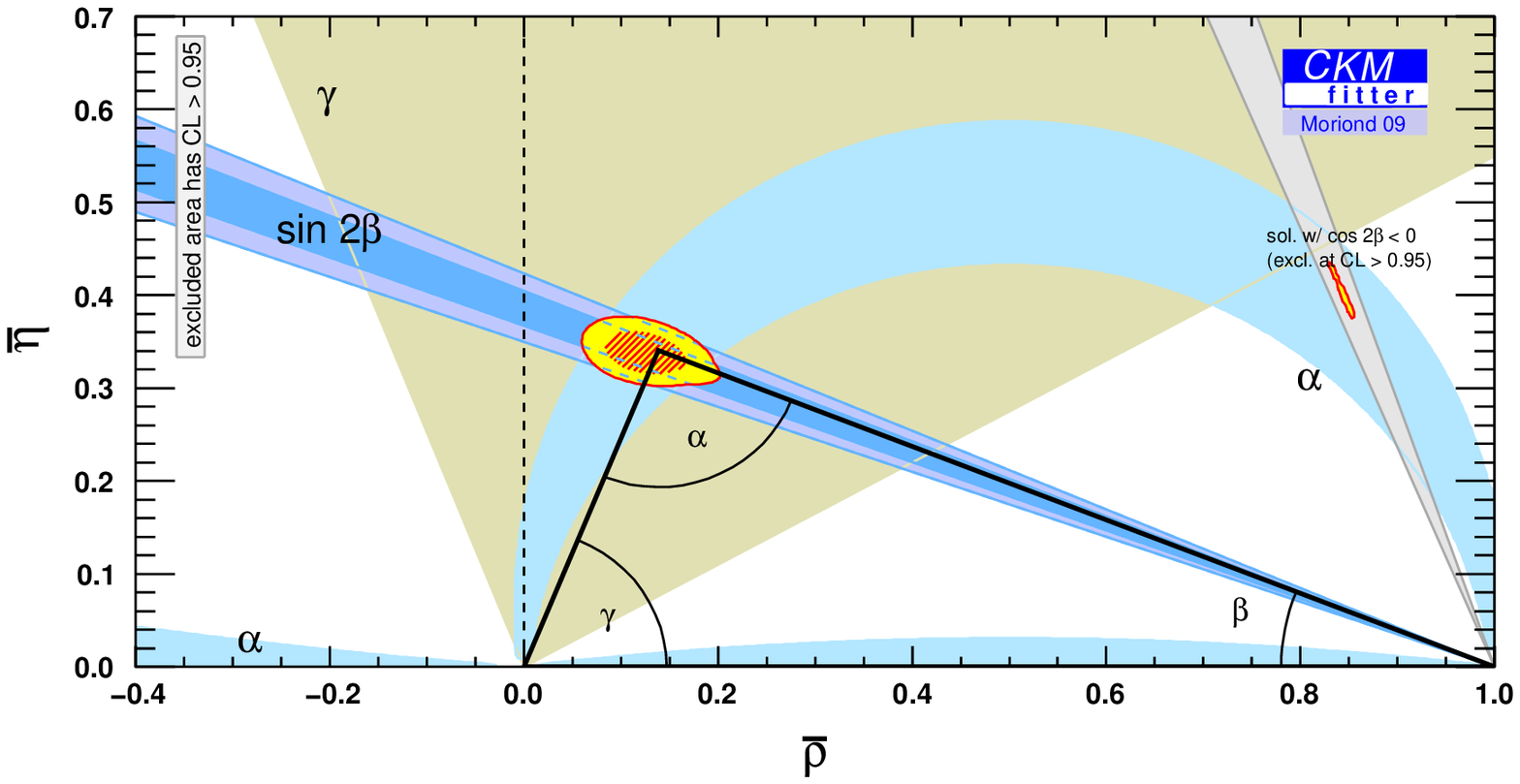}
~~~~~~\includegraphics[width=86mm]{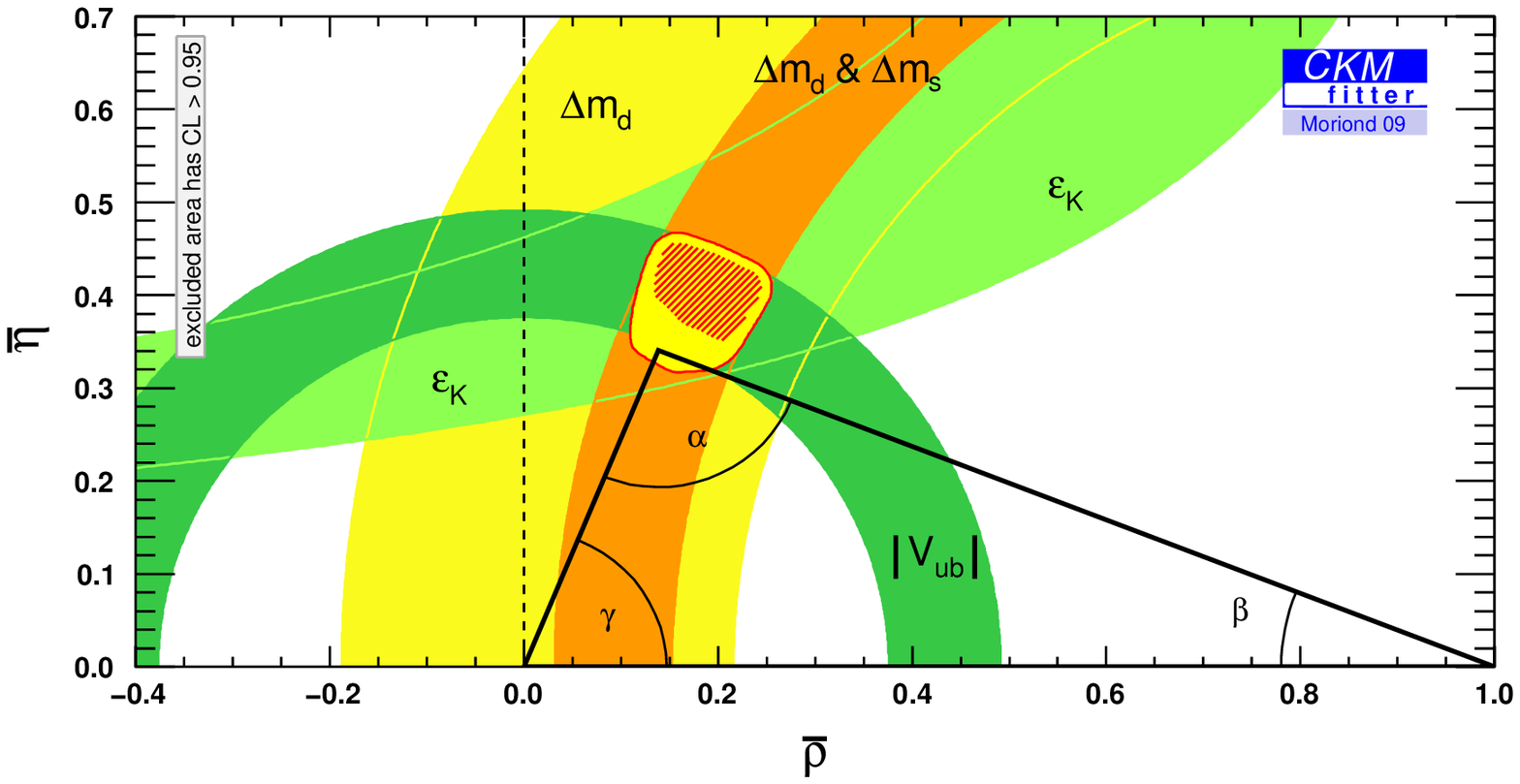}
\caption{ \label{rho_eta}Latest CKMfitter fit of data on the
  $\rho-\eta$ plane describing the unitarity
  triangle\cite{CKMfitter}. The left panel includes only measurements
  of CPV angles, while the right panel excludes them. The green ring,
  $|V_{ub}|$, is dominates by the inclusive semileptonic
  determination.}
\end{figure*}

\subsection{Inclusive}

This has been the method of choice until recently, since it was
thought that the perturbative calculation was reliable and systematic
and hence could be made sufficiently accurate. However it has become
increasingly clear of late that the calculation cannot be made
arbitrarily precise. The method uses effective field theories to
expand the amplitude systematically in inverse powers of a large
energy, either the heavy mass or the energy of the up-quark (or
equivalently, of the hadronic final state). One shows that in the
restricted kinematic region needed for experiment (to enhance the
up-signal to charm-background) the inclusive amplitude is governed by
a non-perturbative ``shape function,'' which is, however, universal:
it also determines other processes, like the radiative $B\to X_s\gamma$. So
the strategy has been to eliminate this unknown, non-perturbative
function from the rates for semileptonic and radiative decays.

Surprisingly, most analysis do not
eliminate the shape function dependence between the two processes.
 Instead, practitioners
commonly use parametrized fits that unavoidably introduce
uncontrolled errors. It is not surprising that errors quoted in the
determination of $|V_{ub}|$ are smaller if
by a parametrized fit than by the elimination method of
\cite{Leibovich:1999xf}. The problem is that
parametrized fits introduce systematic errors that are unaccounted for.

Parametrized fits aside, there is an intrinsic problem with the
method. Universality is violated by sub-leading terms\cite{brickwall}
in the large energy expansion (``sub-leading shape functions''). One
can estimate this uncontrolled correction to be of order
$\alpha_s\Lambda/m_b$, where $\Lambda$ is hadronic scale that
characterizes the sub-leading effects (in the effective theory
language: matrix elements of higher dimension operators). We can try
to estimate these effects using models of sub-leading shape functions
but then one introduces uncontrolled errors into the determination. At
best one should use models to estimate the errors. I think it is fair,
albeit unpopular, to say that this method is limited to a precision of
about 15\%: since there are about 10 sub-leading shape functions, I
estimate the precision as $\sqrt{10}\,\alpha_s\Lambda/m_b$. This is
much larger than the error commonly quoted in the determination of
$|V_{ub}|$.

This is just as well, since the value of $|V_{ub}|$ from inclusives is
in disagreement not only with the value from exclusives but also with
the global unitarity triangle fit.  You can quantify this if you like,
but it is graphically obvious from Fig.~\ref{rho_eta}. The location of
the apex of the unitarity triangle differs in the two panels, and the
agreement would be much better if the green ring, whose radius is
given by $|V_{ub}/V_{cb}|$ and is dominated in the fit by the
determination from the inclusive decay, had smaller radius.

\subsection{Exclusives}

The branching fraction $\text{Br}(B\to\pi\ell \nu) $ is
known\cite{Abe:2004zm} to 8\%. A comparable determination of
$|V_{ub}|$ requires knowledge of the $B\to\pi$ form factor $f_+(q^2)$ to
4\%. There are some things we do know about $f_+$: (i)The shape is
constrained by dispersion relations\cite{Boyd:1994tt}. This means that if we
know $f_+$ at a few well spaced points we can pretty much determine
the whole function $f_+$. (ii)We can get a rough measurement of the
form factor at $q^2=m_\pi^2$ from the rate for $B\to
\pi\pi$\cite{Bauer:2004tj}. This requires a sophisticated effective theory
(SCET) analysis which both shows that the leading order contains a
term with $f_+(m_\pi^2)$ and systematically characterizes the
corrections to the lowest order SCET.  It is safe to assume that
this determination of $f_+(m_\pi^2)$ will not improve beyond the 10\%
mark.

Lattice QCD can determine the form factor, at least over a limited
region of large $q^2$. The experimental and lattice measurements can
be combined using constraints from dispersion relations and
unitarity\cite{Arnesen:2005ez}. Because these constraints follow from
fundamentals, they do not introduce additional uncertainties.  They
improve the determination of $|V_{ub}|$ significantly. The lattice
determination is for the $q^2$-region where the rate is smallest. This
is true even if the form factor is largest there, because in that
region the rate is phase space suppressed. But a rough shape of the
spectrum is experimentally observed, through a binned
measurement\cite{Abe:2004zm}, and the dispersion relation constraints
allows one to combine the full experimental spectrum with the
restricted-$q^2$ lattice measurement.  The best lattice calculations
are in good agreement; however use the same MILC
ensembles\cite{Dalgic:2006dt}. They give 3.55(25)(50) and 3.38(36) for
$10^3|V_{ub}|$. The 11\% error in $|V_{ub}|$ is  dominated by
 lattice errors.

\subsection{Alternatives}
\label{alternatives}
Exclusive and inclusive determinations of
$|V_{ub}|$ have comparable precisions. Neither is very good and the
prospect for significant improvement is limited.  Other methods need
be explored, if not to improve on existing $|V_{ub}|$ to lend
confidence to the result. A lattice-free method would be preferable.
A third method, proposed a while ago\cite{Ligeti:1995yz}, uses the
idea of double ratios\cite{Grinstein:1993ys} to reduce hadronic
uncertainties. Two independent approximate symmetries protect double
ratios from deviations from unity, which are therefore of the order of
the product of two small symmetry breaking parameters. For example,
the double ratio
$(f_{B_s}/f_{B_d})/(f_{D_s}/f_{D_d})=(f_{B_s}/f_{D_s})/(f_{B_d}/f_{D_d})=1+{\cal
O}(m_s/m_c)$ because $f_{B_s}/f_{B_d}=f_{D_s}/f_{D_d}=1$ by $SU(3)$
flavor, while $f_{B_s}/f_{D_s}=f_{B_d}/f_{D_d}=\sqrt{m_c/m_b}$ by
heavy flavor symmetry.  One can extract $|V_{ub}/V_{ts}V_{tb}|$ by
measuring the ratio,
\begin{equation}
\frac{{\rm d}\Gamma(\bar B_d\to\rho\ell\nu)/{\rm d}q^2}{{\rm d}\Gamma(\bar B_d\to K^*\ell^+\ell^-)/{\rm d}q^2}
=\frac{|V_{ub}|^2}{|V_{ts}V_{tb}|^2}\cdot\frac{8\pi^2}{\alpha^2}\cdot\frac1{N(q^2)}\cdot
R_B,
\end{equation}
where $q^2$ is the lepton pair invariant mass, and $ N(q^2)$ is a
known function\cite{Grinstein:2004vb}. When expressed as
functions of the rapidity of the vector meson, $y=E_V/m_V$, the ratios
of helicity amplitudes
\begin{equation}
R_B=\frac{\sum_\lambda |H^{B\to\rho}_\lambda(y)|^2}{\sum_\lambda
  |H^{B\to K^*}_\lambda(y)|^2},\quad
R_D=\frac{\sum_\lambda |H^{D\to\rho}_\lambda(y)|^2}{\sum_\lambda |H^{D\to K^*}_\lambda(y)|^2},
\end{equation}
are related by a double ratio: $R_B(y)=R_D(y)(1+{\cal
  O}(m_s(m_c^{-1}-m_b^{-1})))$. 
This measurement could be done today:  CLEO has accurately measured the
  required semileptonic $D$ decays\cite{Adam:2007pv,Gray:2007pw}.

\begin{figure}
\hspace{-1cm}\includegraphics[width=86mm]{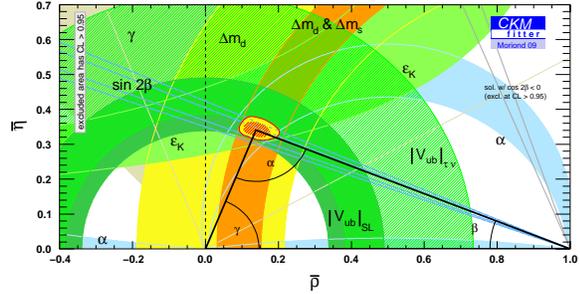}%
\caption{\label{fig:globalCKMfitter} 
  CKMfitter constraints in the $\rho-\eta$ plane including the most
  recent inputs in the global CKM fit\cite{CKMfitter}. The $|V_{ub}|$
  constraint has been split in  two contributions: from
  inclusive and exclusive semileptonic decays (plain dark green) and
   from $B^+\to\tau\nu$ (hashed green). The red hashed region of the
  global combination corresponds to 68\% CL.
}
\end{figure}

More methods are available if we are willing to use rarer decays. To
extract $|V_{ub}|$ from $\text{Br}(B^+\to\tau^+\nu_\tau)
=(1.8\pm0.6)\times10^{-4}$\cite{Mazur:2009zz} one needs a lattice
determination of $f_B$. This was discussed in Sec.~\ref{btotaunu}:
although the determination is still imprecise and relies on lattice
hadronic parameters, it gives an even larger value for $|V_{ub}|$; see
Fig.~\ref{fig:globalCKMfitter} which, however, uses as input
$\text{Br}(B^+\to\tau^+\nu_\tau) = (1.73\pm0.35)\times10^{-4}$, with
substantially lower errors than in the combined BaBar-Belle result of
Ref.~\cite{Mazur:2009zz}. Since we want to move away from relying on
non-perturbative methods (lattice) to extract $|V_{ub}|$ we have
proposed a cleaner but more difficult measurement, the double ratio
\begin{equation}
\frac{\frac{\Gamma(B_u\to\tau\nu)}{\Gamma(B_s\to\ell^+\ell^-)}}{\frac{\Gamma(D_d\to\ell
    \nu)}{\Gamma(D_s\to\ell\nu)}}\sim
\frac{|V_{ub}|^2}{|V_{ts}V_{tb}|^2}\cdot\frac{\pi^2}{\alpha^2}\cdot\left(\frac{f_B/f_{B_s}}{f_D/f_{D_s}}\right)^2.
\end{equation}
In the SM $\text{Br}(B_s\to\mu^+\mu^-)\approx 3.5\times10^{-9}$ $\times
 (f_{B_s}/210\,\text{MeV})^2(|V_{ts}|/0.040)^2$
is the only presently unknown quantity in the double ratio and is
expected to be well measured at the LHC\cite{Schopper:2006he}. 

The ratio $\Gamma(B^+\to\tau^+\nu)/ \Gamma(B_d\to\mu^+\mu^-)$ gives us a fifth method. It
has basically no hadronic uncertainty, since the hadronic factor
$f_B/f_{B_d}=1$, by isospin. It involves$|V_{ub}|^2/|V_{td}V_{tb}|^2$,
an unusual combination of CKMs. In the $\rho-\eta$ plane it forms a circle
centered at $\sim (-0.2,0)$ of radius $\sim0.5$. Of course, measuring
$\Gamma(B_d\to\mu^+\mu^-)$ is extremely hard. 

In a sixth method one
studies wrong charm decays $\bar B_{d,s}\to\bar DX$ (really $b\bar q\to u
\bar c$). This can be done both in semi-inclusive
decays\cite{Falk:1999sa} (an experimentally challenging
measurement) or in exclusive decays\cite{Evans:1999wx} (where
an interesting connection to $B_{d,s}$ mixing matrix elements is
involved).

\section{Rare Radiative $\mathbf{B}$ decays.}
The rare decays $B\to X_s\gamma$ and $B\to X_s\ell^+\ell^-$ are flavor
changing neutral processes that occur first at one loop in the SM. As
such they are sensitive probes of NP. They are complimentary to other
FCNCs and put stringent bounds on NP models beyond  what is obtained
from neutral meson mixing measurements.  Here I will focus on the
radiative decay, $B\to X_s\gamma$ both because of its higher rate and
because of recent progress in theory. Both the total rate and CP
violating asymmetries can probe NP. The average of experimental measurement of
the rate is rather precise\cite{HFAG-BtoXsgamma},
\begin{equation}
\label{eq:btosgammaexp}
\text{Br}(B\to X_s\gamma)^{\text{exp}}_{E_\gamma>1.6~\text{GeV}}
=(355\pm24^{+9}_{-10}\pm3)\times10^{-6}.
\end{equation}
Here we have indicated that the measured rate is only for energetic
photons, $E_\gamma>1.6$~GeV. The combination of data requires some
mild extrapolation since  measurements have differing photon
energy cuts.   BaBar
recently reported a rather strong constraint in the 
CPV asymmetry, $ −0.033 < A_{\text{CP}}(B\to K^∗\gamma) <
0.028$
\cite{:2009we}. 

The theory of $B\to X_s\gamma$ has two parts. The first one is the
computation of the low energy effective Hamiltonian. This is
necessary in order to re-sum large logarithms, $\ln(m_t/m_b)$ or
$\ln(M_W/m_b$,  in the perturbative expansion. The second step is the
computation of the rate from this effective Hamiltonian. 

The effective Hamiltonian, to lowest order in an expansion in $G_F$,
is ${\cal H} = - (4G_F/\sqrt{2}) V^*_{ts}V^{\phantom{*}}_{tb}
\sum_{i=1}^8C_i(\mu)Q_i$. The  $Q_i$ are dimension 6 $\Delta
B=-\Delta S=1$ operators. Roughly they are the tree level four quark
operator and the one it mixes pronouncedly with, $Q_{1}=(\bar
s_L\gamma_\mu c_L)(\bar c_L \gamma^\mu  b_L)$, $Q_{2}=(\bar
s_L\gamma_\mu T^a c_L)(\bar c_L \gamma^\mu T^a b_L)$, four penguin operators, $Q_{3-6}=(\bar s \Gamma_a
b)\sum_{q}(\bar q \Gamma_a' q)$ (with $\Gamma$ matrices in color and
spinor space), and two transition magnetic moment operators
$Q_7=(em_b/16\pi^2)\bar s_L \sigma^{\mu\nu} b_R F_{\mu\nu}$ and
$Q_8=(g_sm_b/16\pi^2)\bar s_L \sigma^{\mu\nu}T^a b_R
G^a_{\mu\nu}$. The problem is to compute reliably the coefficients at
a low renormalization scale, $\mu\sim m_b$. This requires computation
of the coefficients at a short distance scale, $\mu\sim M_W$, and then
using the renormalization group to ``run'' the coefficients to the
physical scale, $\mu\sim m_b$, for which the anomalous dimensions of
the operators needs to be computed. The leading logarithms (LO) were
first summed 20 years ago\cite{Grinstein:1987vj,
  Grinstein:1990tj}. The result is a correction of more than 30\% to
the un-resummed coefficient $C_7(m_b)$ (and therefore a whopping 60\%
effect in the rate). To achieve accuracy comparable with present
experimental measurement it is important to re-sum the next-to-leading
(NLO) and the next-to-NLO (NNLO) logs. This is a challenging
enterprise that has taken the better part of two decades. The NNLO
calculation requires two loop matching of
$C_{1-6}$\cite{Bobeth:1999mk}, three loop matching of
$C_{7-8}$\cite{Misiak:2004ew}, three-loop calculation of the
$(1-6)\times(1-6)$ and $(7-8)\times(7-8)$ blocks of the anomalous
dimension matrix\cite{Gambino:2003zm}, and four-loop of the
$(1-6)\times(7-8)$ block\cite{Czakon:2006ss}.  The magnitude of the
NNLO coefficients are, roughly, $|C_{1,2}(m_b)|\sim1$,
$|C_{3,4,5,6}(m_b)|<0.07$, $C_7(m_b)\simeq-0.3$ and
$C_8(m_b)\simeq-0.15$.

The second step, the computation of the rate from this effective
Hamiltonian is no smaller a feat. To match the accuracy of the
NNLO coefficients one needs a perturbative calculation of the matrix
elements of $Q_{1-6}$ to three-loops\cite{Bieri:2003ue} and of
$Q_{7-8}$ to
two-loops\cite{Blokland:2005uk,Melnikov:2005bx,Asatrian:2006ph}. The
calculation is not complete, some interference terms are missing; see
Ref.~\cite{Misiak:2008ss} for a detailed account and for an account of
non-perturbative effects.

The SM prediction for the branching fraction restricted to energetic
photons $E_\gamma>1.6$~GeV is\cite{Misiak:2006zs}
\begin{equation}
  \text{Br}(B\to X_s\gamma)^{\text{SM}}_{E_\gamma>1.6~\text{GeV}}
  =(3.15\pm0.23)\times10^{-4}
\end{equation}
in magnificent agreement with the experimental average
Eq.~\eqref{eq:btosgammaexp}.

\begin{figure}
\includegraphics[width=80mm]{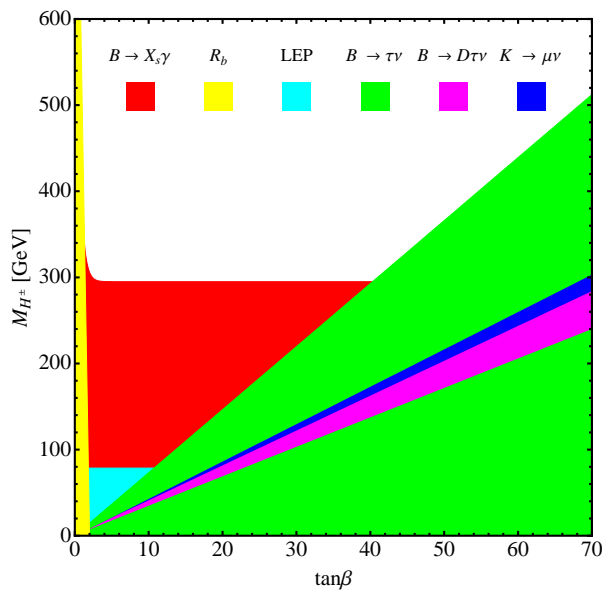}
\caption{\label{fig:haisch-thdm} 
Direct and indirect bounds on $M_{H^\pm}$ in the 
 Two Higgs Doublet Model, type II,  as a function of $\tan\beta$, from
 Ref.~\cite{Haisch:2008ar}.  The shaded 
areas are excluded by the constraints at 95\% CL.}
\end{figure}

The effectiveness of this process in limiting models of NP is nicely
illustrated  in Fig.~\ref{fig:haisch-thdm} which shows the constraints
form a variety of measurements on a two higgs doublet model of type II
(charge-$2/3$ quarks get masses from the vacuum expectation value (VEV) of one
higgs, charge-$-1/3$ quarks form the VEV of the other higgs)  in
the $(M_{H^\pm},\tan\beta)$ plane.  Several measurement nicely
compliment each other, but it is clear that radiative $B$ decays plays
a leading role in excluding parameter space.

We close with a couple of remarks on $B\to X\ell\ell$. The NNLO
calculation in the SM requires one less loop than $B\to X\gamma$ so it
has been complete for quite some time. Much attention has been given
to the observation that the forward-backward asymmetry in $B\to
K^*\ell\ell$ has a zero in the SM\cite{Burdman:1995ks}. The presence
and location of the zero suffer little from hadronic uncertainties and
from contamination from non-resonant $B\to K\pi\ell\ell$
decays\cite{Grinstein:2005ud}. Less well known is the fact that at
large invariant lepton-pair mass $B\to K^*\ell\ell$ is well understood
and that tests of the SM can be done, largely free of hadronic
uncertainties, by a method of double ratios\cite{Grinstein:2004vb}
as described in Sec.~\ref{alternatives}. Unfortunately
experimentalists have not conducted this test, even if the data is
available.

\begin{acknowledgments}
Work supported in part by the US Department of Energy under contract
DE-FG03-97ER40546.
\end{acknowledgments}

\bigskip 

\end{document}